\documentclass[twoside,twocolumn,9pt]{article}
\usepackage{extsizes}
\usepackage[super,sort&compress,comma]{natbib} 
\usepackage[version=3]{mhchem}
\usepackage[left=1.5cm, right=1.5cm, top=1.785cm, bottom=2.0cm]{geometry}
\usepackage{balance}
\usepackage{times,mathptmx}
\usepackage{sectsty}
\usepackage{graphicx} 
\usepackage{lastpage}
\usepackage[format=plain,justification=justified,singlelinecheck=false,font={stretch=1.125,small,sf},labelfont=bf,labelsep=space]{caption}
\usepackage{float}
\usepackage{fancyhdr}
\usepackage{fnpos}
\usepackage[english]{babel}
\usepackage{array}
\usepackage{droidsans}
\usepackage{charter}
\usepackage[T1]{fontenc}
\usepackage[usenames,dvipsnames]{xcolor}
\usepackage{setspace}
\usepackage[compact]{titlesec}
\usepackage{hyperref}

\definecolor{cream}{RGB}{222,217,201}

\begin{document}

\pagestyle{fancy}
\thispagestyle{plain}
\fancypagestyle{plain}{
%
}

\makeFNbottom
\makeatletter
\renewcommand\LARGE{\@setfontsize\LARGE{15pt}{17}}
\renewcommand\Large{\@setfontsize\Large{12pt}{14}}
\renewcommand\large{\@setfontsize\large{10pt}{12}}
\renewcommand\footnotesize{\@setfontsize\footnotesize{7pt}{10}}
\makeatother

\renewcommand{\thefootnote}{\fnsymbol{footnote}}
\renewcommand\footnoterule{\vspace*{1pt}%
\color{black}\hrule width 3.5in height 0.4pt \color{black}\vspace*{5pt}} 
\setcounter{secnumdepth}{5}

\makeatletter 
\renewcommand\@biblabel[1]{#1}            
\renewcommand\@makefntext[1]%
{\noindent\makebox[0pt][r]{\@thefnmark\,}#1}
\makeatother 
\renewcommand{\figurename}{\small{Fig.}~}
\sectionfont{\sffamily\Large}
\subsectionfont{\normalsize}
\subsubsectionfont{\bf}
\setstretch{1.125} 
\setlength{\skip\footins}{0.8cm}
\setlength{\footnotesep}{0.25cm}
\setlength{\jot}{10pt}
\titlespacing*{\section}{0pt}{4pt}{4pt}
\titlespacing*{\subsection}{0pt}{15pt}{1pt}

\fancyfoot[LE]{\footnotesize{\sffamily{\thepage~\textbar\hspace{3.45cm} 1--\pageref{LastPage}}}}
\fancyhead{}
\renewcommand{\headrulewidth}{0pt} 
\renewcommand{\footrulewidth}{0pt}
\setlength{\arrayrulewidth}{1pt}
\setlength{\columnsep}{6.5mm}
\setlength\bibsep{1pt}

\makeatletter 
\newlength{\figrulesep} 
\setlength{\figrulesep}{0.5\textfloatsep} 

\newcommand{\topfigrule}{\vspace*{-1pt}%
\noindent{\color{black}\rule[-\figrulesep]{\columnwidth}{0.5pt}} }

\newcommand{\botfigrule}{\vspace*{-2pt}%
\noindent{\color{black}\rule[\figrulesep]{\columnwidth}{0.5pt}} }

\newcommand{\dblfigrule}{\vspace*{-1pt}%
\noindent{\color{black}\rule[-\figrulesep]{\textwidth}{0.5pt}} }

\makeatother

\twocolumn[
\vspace{3cm}
\sffamily

\LARGE{\textbf{Coalescence driven self-organization of growing \mbox{nanodroplets} around a microcap}} \\

\noindent\large{Brendan Dyett,\textit{$^{a}$} Hao Hao,\textit{$^{b}$} Detlef Lohse\textit{$^{c}$} and Xuehua Zhang\textit{$^{d,a,c*}$}} \\

\noindent\normalsize{The coalescence between growing droplets is important for the surface coverage and spatial arrangements of droplets on surfaces. In this work, total internal reflection fluorescence (TIRF) microscopy is utilized to \textit{in-situ} investigate the formation of nanodroplets around the rim of a polymer microcap, with sub-micron spatial and millisecond temporal resolution. We observe that the coalescence among droplets occurs frequently during their growth by solvent exchange. 	Our experimental results show that the position of the droplet from two merged droplets is related to the size of the parent droplets. The position of the coalesced droplet and the ratio of parent droplet sizes obey a scaling law, reflecting a coalescence preference based on the size inequality. As a result of droplet coalescence,  the angles between the centroids of two neighbouring droplets increase with time, obeying a nearly symmetrical arrangement of droplets at various time intervals. The evolution of the position and number from coalescence of growing droplets  is modelled. The mechanism for coalescence driven self-organization of growing droplets is general, applicable to microcaps of different sizes and droplets of different liquids.  The understanding from this work may be valuable for positioning nanodroplets by nucleation and growth without using templates. 
} 



\vspace{0.6cm}

]  


\renewcommand*\rmdefault{bch}\normalfont\upshape
\rmfamily
\section*{}
\vspace{-1cm}


\footnotetext{\textit{$^{a}$Soft Matter $\&$ Interfaces Group, School of Engineering, RMIT University, Melbourne, VIC 3001}}
\footnotetext{\textit{$^{b}$Electrical and Biomedical Engineering, RMIT University, Melbourne, VIC 3001, Australia.}}
\footnotetext{\textit{$^{c}$Physics of Fluids group, Department of Science and Engineering, Mesa+ Institute, J. M. Burgers Centre for Fluid Dynamics and the Max Planck Center Twente for Complex Fluid Dynamics, University of Twente, P.O. Box 217, 7500 AE Enschede, The Netherlands}}
\footnotetext{\textit{$^{d}$Department of Chemical \& Materials Engineering, Faculty of Engineering, University of Alberta,
Edmonton, Alberta T6G1H9, Canada}}
\footnotetext{\textit{E-mail: xuehua.zhang@ualberta.ca}}

\footnotetext{\dag~Electronic Supplementary Information (ESI) available: See DOI: 10.1039/b000000x/}








\section{Introduction}
The coalescence of droplets has drawn intensive research interest due to its importance in many fundamental and applied processes\cite{dewformation,ripening_correlation,ripening_twodimensions}.  For example, controlling coalescence is essential for material templating from droplets\cite{Bunz2006,ZhangLohse2015}, while compartmentalized reactions within microfluidic systems may be triggered by coalescence\cite{song2006reactions}. The physical phenomenon of jumping droplets from coalescence on low adhesive substrates has also gained significant interest with potential applications in water collection, self-cleaning and anti-icing surfaces, condensation heat transfer, energy harvesting and power generation\cite{heattransfer, superhydrophobic_heattransfer,watercollectionbeetle,jumpingdrops, EWang2014}. A great deal of research has been conducted toward understanding the dynamical events which occur during droplet coalescence, in particular relating to the size of two parent droplets and position of the new droplet.  Thoroddsen et al. used high speed imaging to identify the critical diameter ratio, above which a satellite is produced during the coalescence of millimeter-sized drops \cite{thoroddsen2009}. Similarly, a critical size ratio has been determined for inducing jumping droplets as condensate on a superhydrophobic surface\cite{Wang2017}. Rykaczewski et al. studied droplet condensation on a superhydrophobic surface and observed that large droplets would consume smaller satellite droplets with minimal shift in the larger droplet location\cite{Jones2011}. Following this coalescence, the nucleation site of the smaller droplet was free to nucleate additional droplets. 

The mechanism for the position shift from coalescence was later explained by Weon and Je, who demonstrated the position of merged droplets (or bubbles) critically depends on the size ratio of the two parents\cite{Je2012}. As the ratio (large/small) increases, the final position is closer to the larger parent (named `coalescence preference')\cite{Je2012}. The displacement was attributed to surface energy release from the coalescence and followed the scaling law $(displacement_{large}/displacement_{small}) \sim (radius_{large}/radius_{small})^{-p}$. The exponent $p$ varied from 5 for loosely packed microbubbles to 2 for densely packed microbubbles\cite{preference_dense}. Clearly, the coalescence preference is critical in the dynamics of colloids; however the general mechanism is still to be determined\cite{lbm_preference}. 

The coalescence of two droplets comprises the initial formation of a microscopic liquid bridge,\cite{coalescence1999,hernandez2012symmetric,eddi2013influence} which then relaxes as the merged droplet recovers a spherical equilibrium shape. The dynamics of this process can be characterized by the respective Ohnesorge number $Oh$, a dimensionless number which describes the relative viscous to inertial and surface forces,  given by $Oh = \mu/\sqrt{\rho \sigma r}$ where $\mu$, $\rho$, $\sigma$ are the dynamical viscosity, density and interfacial tension respectively while $r$ usually denotes the radius of the larger of the coalescing droplets \cite{Farokhirad2015}. 

The study of coalescence on micro-/nano- structured surfaces has typically encompassed drop radii in the range of 2 $\mu$m - 2 mm\cite{Chu2016,Miljkovic2016,Boreyko2017}. Recent work has identified necessary conditions for inertia or viscous forces to be negligible within this size regime \cite{regimes,hernandez2012symmetric,eddi2013influence}. Up to now however, the dynamics of coalescence between submicron droplets has rarely been quantitatively studied. Here, we fill this gap and study the coalescence between surface nanodroplets.


Surface nanodroplets in this study are typically 10-100 nm in height and 0.1-1 $\mu$m in lateral diameter. A simple process to produce these droplets is standard solvent exchange wherein a solution of the droplet liquid in a good solvent is displaced by a poor solvent under controlled flow conditions \cite{Zhang2015, viscous}. The growth rate of surface nanodroplets is determined by an oversaturation pulse created at the mixing front \cite{Zhang2015,lu2017universal}.  The growth of the droplets during the solvent exchange can be maintained at a very low rate such that the expansion velocity of the droplet wall is negligible in the time scale of the coalescence process.  Hence in this situation two droplets meet with very low external velocity at the onset of the coalescence. 

These growing surface nanodroplets are confined to the rim of the microcap structures on the substrate \cite{Peng2016}.  They meet neighbouring droplets with freedom only in azimuthal direction around the rim due to pinning, eliminating uncontrolled coalescence of multiple droplets in various directions.  Our recent work showed that these droplets confined to the rim of the microcap form remarkable symmetric arrangements during the solvent exchange \cite{Peng2015}. It is clear that the droplet interactions are crucial for their arrangement, however, the exact mechanism controlling the pattern formation remains unclear without visualizing the process of the droplet growth and position with time\cite{Peng2015}. To date, the growing process of these nanodroplets had not yet been experimentally monitored because of strong light scattering from droplets in the bulk media.

A method to overcome this difficulty is total internal reflection fluorescence microscopy (TIRF), which has recently been utilized to characterize the dynamics of surface nanobubbles.\cite{Chan2012, Ohl2017} By virtue of an evanescent wave, TIRF only illuminates nanometers above the substrate.\cite{tirf1989} The selective illumination negates background noise from the droplets within the bulk and overcomes contrast issues due to media thickness. As a consequence, TIRF provides sufficient spatial resolution to distinguish sub-micron surface droplets on a substrate in contact with a scattering liquid phase. With sufficient intensity, the droplet growth can be captured with a framerate of about 30 - 100 fps. This combination of temporal and spatial resolution allows for capturing the diffusive growth and coalescence of submicron droplets \textit{in-situ}.  

In this paper, we apply TIRF for the first time to follow the growth and self-organisation of nanodroplets during the solvent exchange.  The $Oh$ number varies from 1 to 3 for different droplet sizes, indicating slight dominance of the viscous over capillary forces.
With the TIRF method we are able to elucidate the mechanism for the self-organisation of the surface nanodroplets around the rim of a microcap. The understanding from this work may be valuable for control of nanodroplet spatial arrangement during heterogeneous nucleation.

\section {Experimental methods}

\begin{figure*}[!htp]	
	\includegraphics[trim={0 13.5cm 0 0}, clip, width=2\columnwidth]{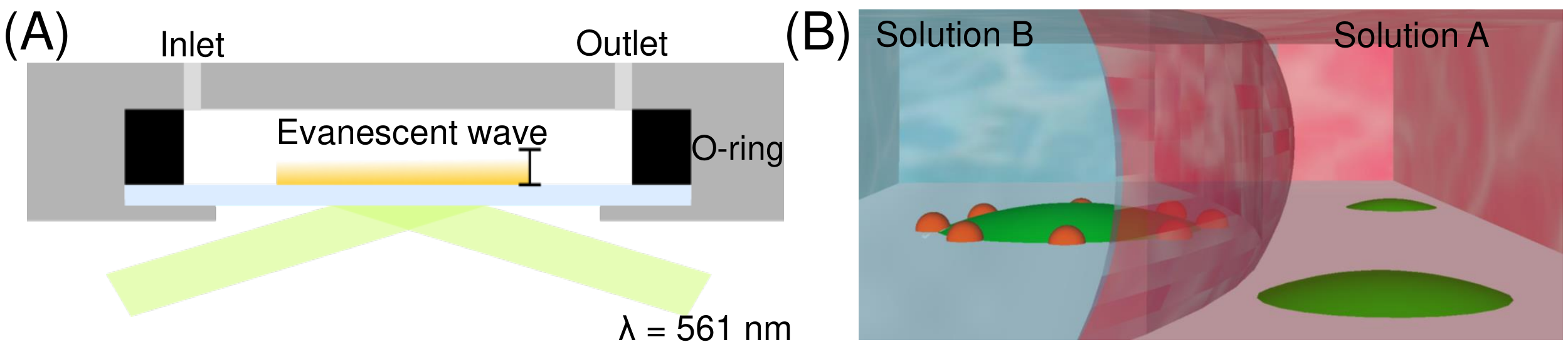}
	\centering
	\caption{(A) Schematics of the fluid cell for the {\it in-situ} observation of droplet formation through solvent exchange. Surface droplets are formed on the glass substrate. Owing to the evanescent wave from total internal reflection, only droplets near the substrate are visualized. (B)  Schematics of droplet growth around a microcap $(green)$ during solvent exchange. The right side of the schematic shows oil-rich solution A $(red)$, which is displaced by the water-rich solution B $(blue)$. The droplets forming as the mixing front passes are represent by spheres $(red)$.}
	\label{schematic}	
\end{figure*}

\subsection {Chemicals, materials and substrates}
1-Octanol (99\%, Merck) and 1,6-hexanediol diacrylate (HDODA) (80\%, Sigma) were the droplet liquids. Ethanol (AR, Chem-supply) was used as the good solvent.  Octyldecyltrichlorosilane (OTS) (99\%, Sigma) was used to modify the surface wettability. Nile-red (Sigma) and rhodamine 6G (Sigma) were all used in TIRF imaging. All the above chemicals were used as received. Water (Milli-Q, 18 M$\Omega$/cm) was used in all experiments. Glass substrates ($\phi$ 42 mm, No. 1, Proscitech) were used after hydrophobic surface functionalization following a previously reported procedure\cite{sam2014}. The microcap substrate was prepared by photopolymerization of droplets on a smooth substrate, following the solvent exchange protocol reported in several of our previous papers \cite{Peng2015}.   In brief, the fluid cell with the smooth substrate placed inside was charged with a solution of 0.5 \% HDODA in 50 \% ethanol-water, which was gradually displaced by a second solution of HDODA-saturated water. The displacement of the second solution was controlled by a digital syringe pump (NE-1000, Pump Systems).  Following the exchange the fluid cell was placed under UV light (20 W, 365 nm, Thermofisher) to cure the monomer droplets yielding a microcap patterned glass. 

\begin{table}[h]

\small
  \caption{Physical properties of the two droplet liquids in this study.}
  \label{Table1}
  
  \begin{tabular*}{0.48\textwidth}{@{\extracolsep{\fill}}lllll}
  	\hline 
    Oil & $\gamma_{oil-water}$ & Contact angle & $\rho$ & $\mu$ \\
    & $(mNm^{-1})$ & $(^\circ)$ &  $(kgm^{-3})$ & (mPa $\cdot$s)\\
    & & & &\\  
    \hline
    & & & &\\
    1-Octanol & 8.5 & 54 & 824 & 7 \\
    & & & & \\
    HDODA & 35 & 18 & 1010 & 6 \\ 
    \hline
    \end{tabular*}
     
\end{table}

\subsection{TIRF measurements of growing surface nanodroplets}  

To form HDODA droplets on the substrate with microcaps, the standard solvent exchange was performed. For 1-octanol droplets, the first solution was 2\% 1-octanol in 50 \% ethanol aqueous solution, while the second solution was 1-octanol saturated water. In each case the initial and final solution were stained by fluorescent markers, namely by nile-red or rhodamine 6G.

A home-built fluid cell was used for solvent exchange in the measurements by TIRF, as shown in Figure \ref{schematic}A. Two acrylic frames provide structural support for circular glass substrates (42 mm diameter). The base plate was designed such that only the edges are supported, allowing the objective to directly image the base of the glass. The liquid seal and channel height is maintained at 560 $\mu$m by a silicone o-ring.

The growth behaviour of the droplets was followed \textit{in-situ} by TIRF on a Nikon N-Storm super resolution confocal microscope (TIRF 100x 1.49 NA objective cap). The droplet formation is depicted in Figure \ref{schematic}B.  The video was taken by using continuous wave 561 nm laser, at $\sim$ 10 \% intensity with all filters in the out position. Within the NIS-Elements AR software the TIRF mirror position was adjusted until achieving total internal reflection, determined by a simultaneous decrease in background brightness.  The region of interest was collected by an Andor iXon EMCCD camera and defined as 40.96 $\times$ 40.96 $\mu$m$^2$ with $\sim$ 17.46 ms exposure time or 81.92 $\times$ 81.92 $\mu$m$^2$ with $\sim$ 31.20 ms. In each case resulting in a pixel calibration of 0.16 $\mu$m per pixel.  Under TIRF, droplets near the substrate are selectively captured within the submicron regime. In general, TIRF is able to capture droplets from radius of $\sim$ 200 nm and above. In the case of solvent exchange utilizing the microcap patterned substrate, the camera settings described above allow for temporal resolution in the order of milliseconds.

\subsection{Data analysis}  

To analyse the droplet number, size and location as a function of time, it was necessary to process the collected images to distinguish droplets interacting along the microcap rim. The semi-automated image analysis was programmed using Matlab (MathWorks Inc.). Each frame of the raw data was extracted and analysed. A mask was utilized to specify the region of interest. To prepare the mask, a frame is manually identified with a peak number of droplets. From this frame, three droplets out of the many along the microcap rim with an angle around 120$^\circ$ in between them were identified. We then calculate the center and radius of a circle through the centres of three droplets. Based on these results, an inner mask and an outer mask were built to exclude non-related droplets.  Both masks were adjusted with the increasing droplet size. The images were enhanced using unsharp masking and remapping intensity values to eliminate background interference. The droplets were then segmented using fixed threshold after adaptive histogram equalization. Finally, an image blob analysis was utilized to calculate the centroid and area of each droplet from which the angles between neighbouring droplets could be calculated.

\section{Results and Discussion}
\subsection{Coalescence between growing surface nanodroplets}

\begin{figure*}[!ht]

	\includegraphics[trim={0 11.5cm 0 0},clip, width=2\columnwidth]{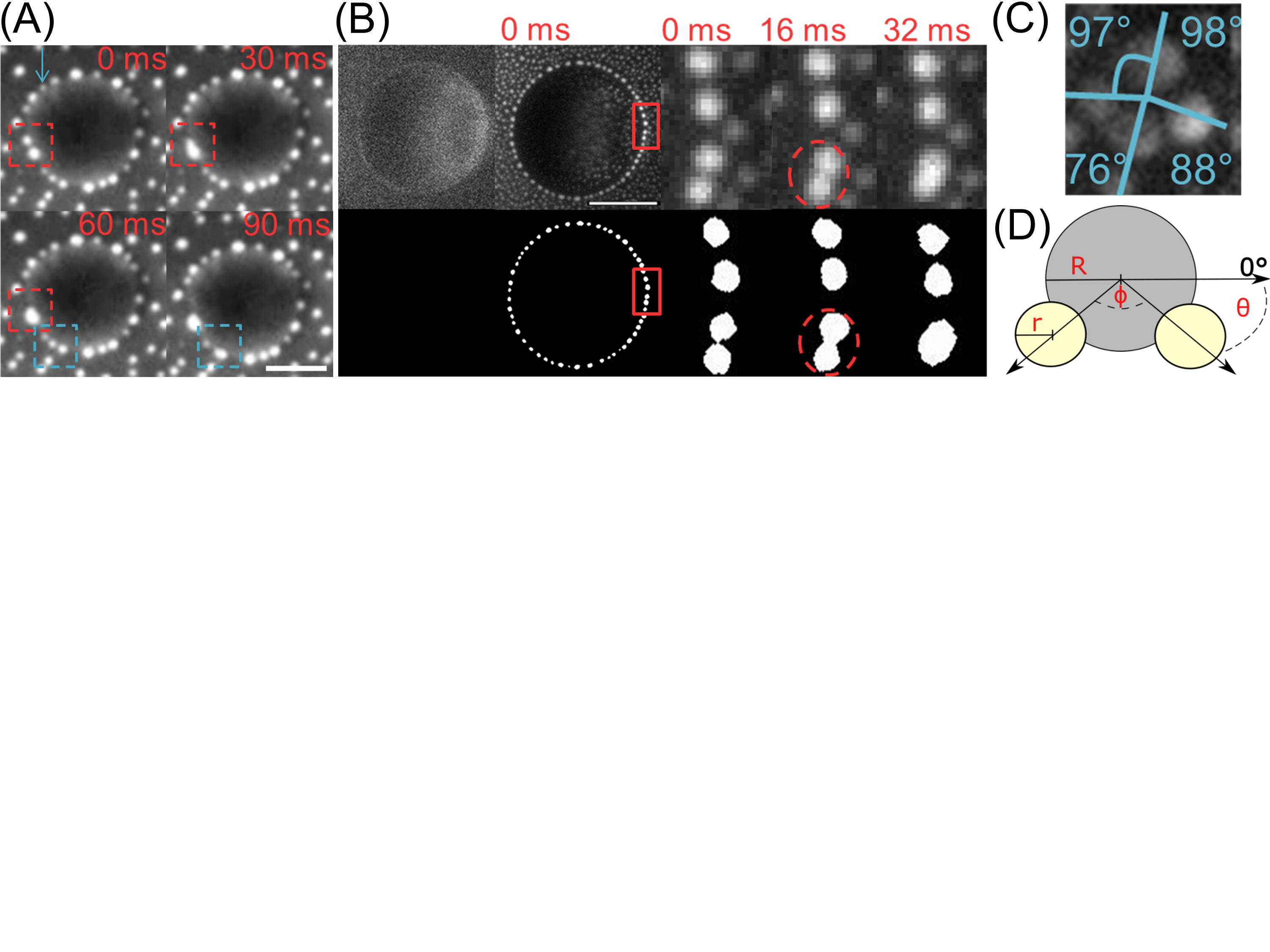}
	\centering
	\caption{(A) Frame-by-frame snap shots of growing nanodroplets formation around a microcap. Each frame represents a 30 ms step. Within the red box from 0 to 60 ms, coalescence between droplets, 682 nm and 970 nm in diameter, is observed, including part of the bridge formation at 30 ms. Within the cyan box, coalescence between the other two droplets, 606 nm and 766 nm in diameter is observed. The smallest visible droplet of 408 nm diameter is indicated by the cyan arrow at 0 ms. Scale bar = 5 $\mu$m. (B) Snapshots showing droplet growth around a 23.2 $\mu$m microcap. The top and bottom row represent raw and processed image respectively. Only droplets within the vicinity of the rim are shown for the processed images. The first two images show the bare lens and the same lens 7.7 seconds after the initial droplet detection. In both series of images, the inset is highlighted by the red box, indicating the detection of droplets merging and coalescing around the cap. These insets show droplet coalescence in 16 ms time steps. The time is the same for both the raw and the processed image series. Scale bar = 10 $\mu$m. (C) Labelled TIRF image of four droplets around a lens with angles of 97$^\circ$, 98$^\circ$, 88$^\circ$ and 76$^\circ$ between two respective droplets. (D) Schematic representation used to describe the droplet growth; `r' and `R' refer to the radius of the droplet and microcap. The absolute anglular position ($\theta$) is defined as the angle away from a horizontal reference plane, which is subsequently used to determine $\phi$, the relative angle between droplets.}
	\label{TIRFDroplets}
\end{figure*}

The time series of droplet growth shown in Figure \ref{TIRFDroplets}A demonstrates the spatial and temporal resolution in our TIRF measurements. The sequential events during the nanodroplet coalescence are resolved in $\sim$30 ms steps.  Two initial droplets with 682 nm and 970 nm respective diameters are observed at 0 ms, a stage of bridging between them at 30 ms, and then the merged droplet at 60 ms.  Droplet coalescence is also observed between two droplets with diameter of 606 nm and 766 nm, respectively, at 90 ms. The smallest droplet diameter with an initial diameter 408 nm is also clearly resolved throughout the series.  Such resolutions allow us to quantitatively characterize the droplet size and location before and after the coalescence. 


Figure \ref{TIRFDroplets}B shows a microcap and growing droplets around it over a longer period of 7$s$. The cap diameter is 23.2 $\mu$m and the cap is surrounded by a necklace of 52 droplets. The droplets around the rim become more visible in the binary processed images.  In the view of the same focal plane, we are able to resolve the coalescence events between droplets at the rim. At approximately 7.7$s$ after initial droplet formation, the coalescence is highlighted in 16 ms steps. During growth each droplet is characterized according to the schematic Figure \ref{TIRFDroplets}D. 

\begin{figure}[!ht]
	\includegraphics[width=1\columnwidth]{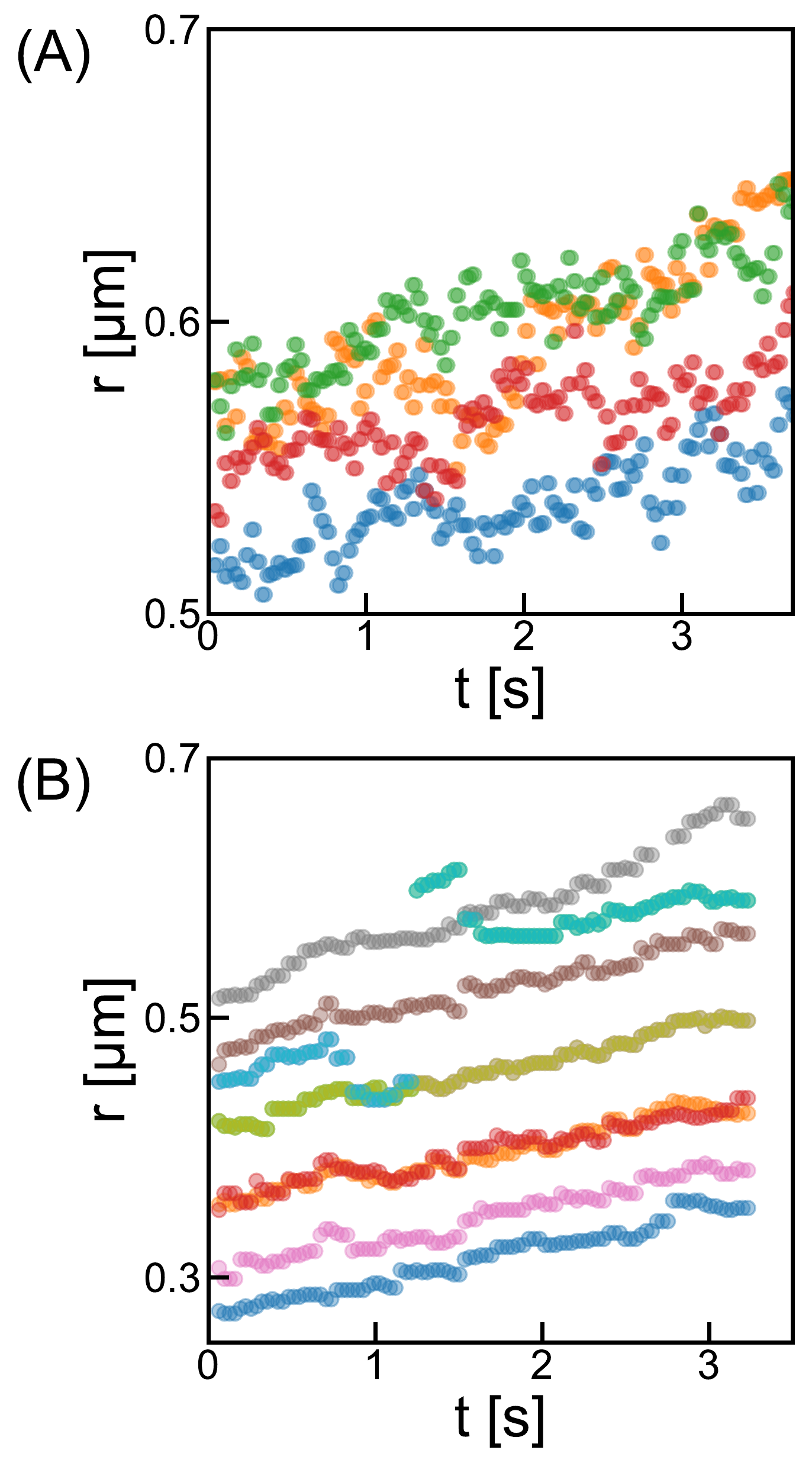}
	\centering
	\caption{(A) Growth of HDODA droplets around a 3.46 $\mu$m microcap given by individual droplet radius vs time. The droplets are differentiated by colour. (B) Growth of 1-octanol droplets around a 8.33 $\mu$m microcap given by individual droplet radius vs time. Here, t = 0 s does not indicate the beginning of the solvent exchange. Rather, t = 0 s corresponds to the beginning of the growth interval which was selected due to the absence of coalescence events.}
	\label{individualgrowth}
	\centering
\end{figure}

The growth of individual droplets is shown in Figure \ref{individualgrowth} and Figure S1. Both figures show the droplet size during selected time intervals during which no coalescence occurred around the microcap. The plots demonstrate that the droplets of different sizes all continually increase. We analysed more than 150 droplets but did not observe any incident where a droplet reduced in size. Moreover, given that droplet coalescence can be influenced by the initial conditions\cite{Nikolayev2004}, these results indicate that the droplet growth (and velocity) is negligible over the relevant time frame for coalescence $\sim$ 16 ms.

Throughout our experiment we observe droplets radii in the range of $\sim$ 200 nm - 1400 nm.  From these radii and the physical properties provided in Table 1, we determine the $Oh$ to be of the order of $\sim$ 1 - 3. Here we consider $r$ to be the larger of the droplet pair, $\mu$ the droplet liquid viscosity, $\rho$ the droplet liquid density and $\sigma$ the droplet-water interfacial tension. 
E.g. For HDODA, $r$ = 1 $\mu$m, $\rho$ = 1010 $kg/m^3$, $\sigma$ = 35 mN/m, $\mu$ = 6 $mPa\cdot s$, yields $Oh$ $\approx$ 1. From these values of $Oh$ it is expected that viscous dissipation hinders any dramatic events such as droplet removal from the surface.\cite{Zhao2011}

\subsection{Position shift of coalescing nanodroplets} 

\begin{figure}[htp]
	\centering
	\includegraphics[width=0.9\columnwidth]{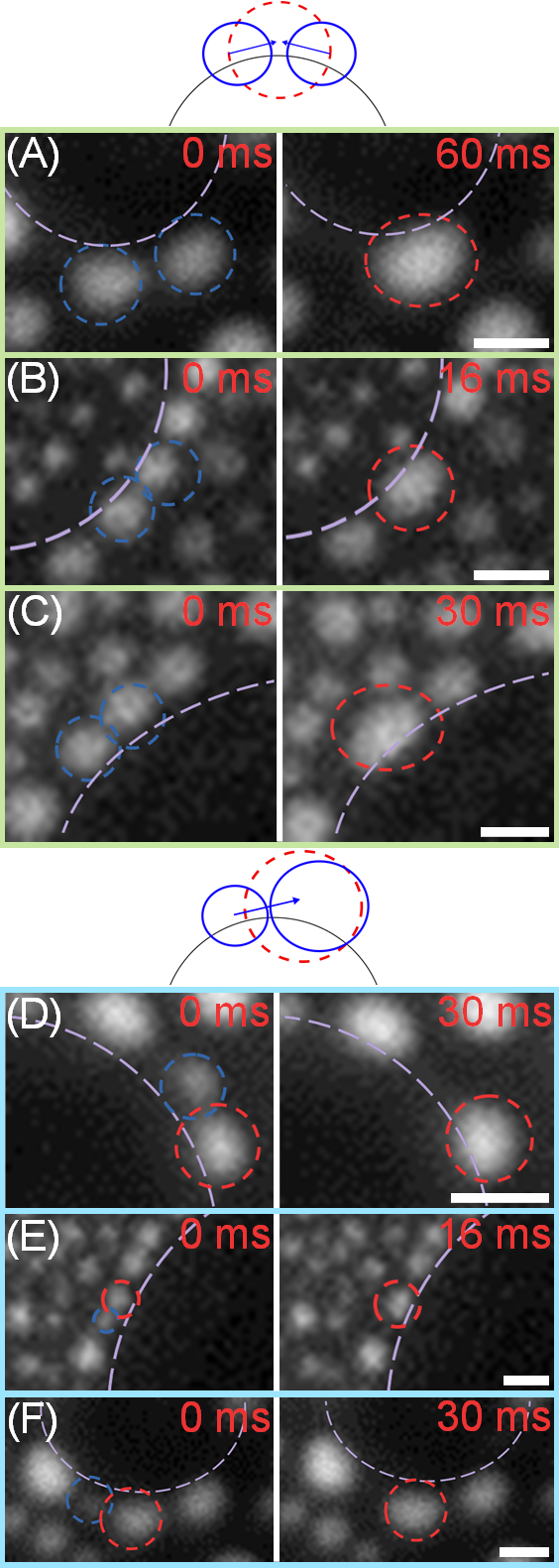}
	\caption{Sketmatics and experimental TIRF images of droplet pairs around the microcap. We show coalescence events between pairs of similar by (A-C) and dissimilar by (D-E) sized droplets, respectively, respectively. Scale bar = 2 $\mu$m.}
	\label{rimrim}
\end{figure}

As depicted within the schematics of Figure \ref{TIRFDroplets}D, the position of a droplet is described by its absolute angle $\theta$, measured against the horizontal reference plane. These angles can then be used to determine the relative angle between the droplets (noted as $\phi$). Figure \ref{rimrim} and Table \ref{coalescencetable} show the detailed coalescence process of two groups of droplets along the microcap rim. In (A)-(C), the two droplets are similar in size before coalescence. The final droplet is located in the middle between the two merged droplets. For example in Figure \ref{rimrim}A, two droplets with radii of 923 nm and 960 nm are initially positioned at 90.1$^\circ$ and 51.8$^\circ$, respectively. After coalescence, the final position is 70.5$^\circ$, indicating a near even shift of 19.6$^\circ$ and 18.7$^\circ$, respectively. 

In Figure Figure \ref{rimrim}(D)-(F), the two droplets are very different in size before coalescence.  It is evident that larger droplets are less mobile than the smaller one when merging with them. For example in Figure \ref{rimrim}F, two droplets with radii of 890 nm and 1321 nm are initially positioned at 118.8$^\circ$ and 97.1$^\circ$, respectively. Following coalescence the final position is 102.5$^\circ$, indicating a shift of 16.3$^\circ$ and 5.4$^\circ$ for the small and large droplet, respectively. As a result the final droplet is positioned closer to the larger of the parent droplets. The results from tracking the nanodroplets with time \textit{in-situ} clearly reveals that there is preference for the droplet coalescence around the rim. 
 
\begin{table}[h]
	
	\small
	\caption{The droplet sizes and angular positions for the droplet coalescence shown in Figure \ref{rimrim}}
	\label{coalescencetable}
	
	\begin{tabular*}{0.48\textwidth}{@{\extracolsep{\fill}}llllll}
		\hline 
		Caption & $r_1$ & $r_2$ & $\theta_1$ & $\theta_2$ & $\theta_f$ \\
		&  $(nm)$ & $(nm)$& ($ ^\circ$)  & ($ ^\circ$)& ($ ^\circ$) \\
		\hline
		(A)	&923 &960 &90.1 &51.8 &	71.5	\\
		(B)	&920  &947  &139.3  &133.4  &135.5	\\
		(C)	&960 &1047 &88.7 &81.5 &84.7		\\
		(D) &360  &846  &54.1  &50.5  &51.0	\\ 
		(E)	&370 &446 &157.6 &154.4 &155.4		\\
		(F)	&890  &1321  &118.8  &97.1  &102.0 	\\
		\hline
	\end{tabular*}
	
\end{table}

\begin{figure}[!ht]
	\includegraphics[trim={0 0 0 0}, clip,width=\columnwidth]{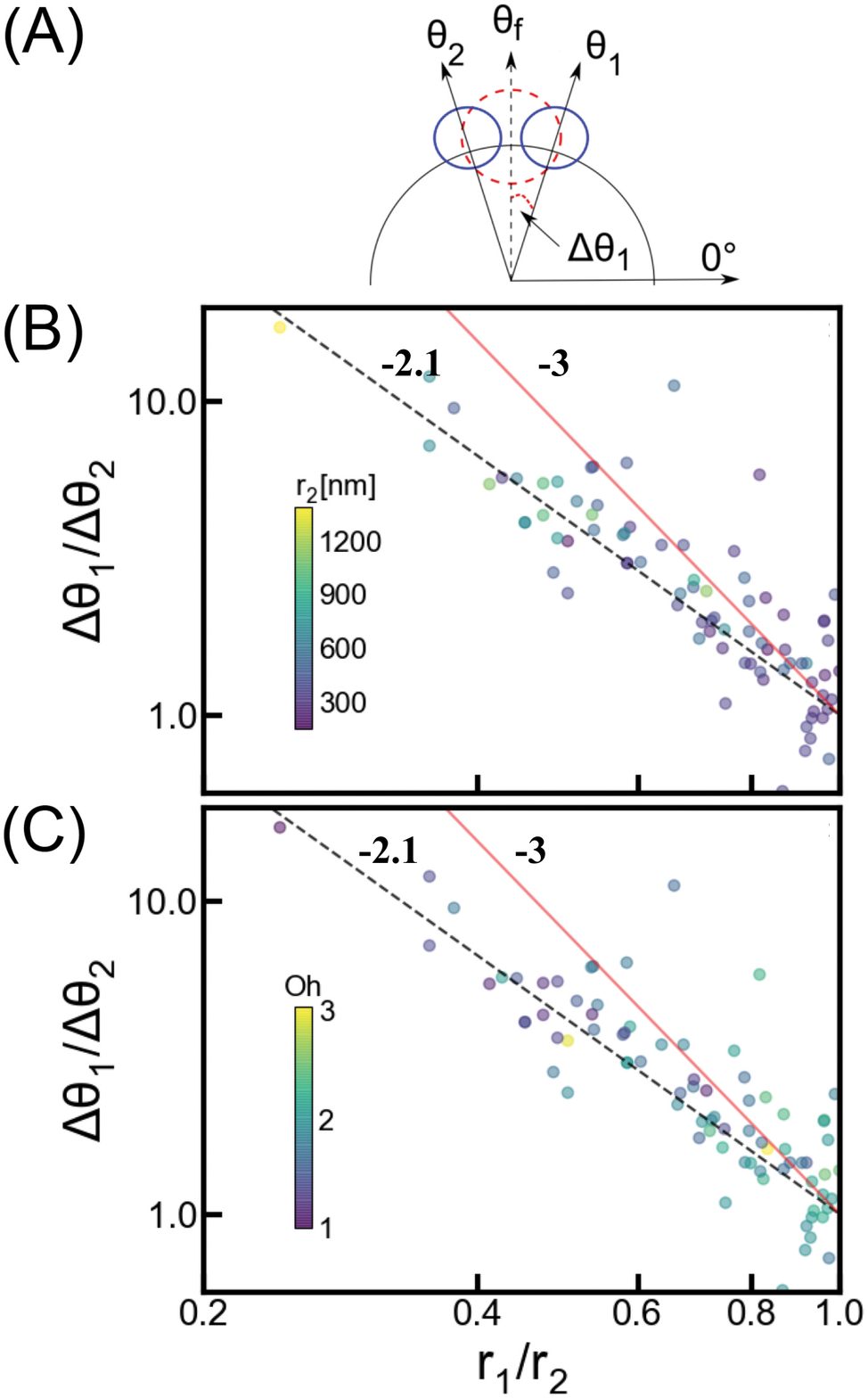}

	\caption{(A) Schematics of coalescence. Parent droplets are indicate by solid blue lines, while the final droplet is indicated by the dashed red line. The initial droplet angles are $\theta_1$ and $\theta_2$. $\Delta \theta_1$ and $\Delta \theta_2$ are defined as their respective difference to the final droplet position, $\theta_f$. Each angle is originally defined against a horizontal reference plane. (B-C) Plot of the ratio of angle shifts relative to droplet sizes, $\frac{\Delta \theta_1}{\Delta \theta_2}$ vs $\frac{r_1}{r_2}$. The experimental data are shown as coloured circles, coloured by $r_2$ respective $Oh$ in B and C. In both cases, the best experimental power law fit, $\Delta\theta_1 / \Delta\theta_2 \sim (r_1/r_2)^{-p}$ with $p$ = 2.1 and the power law from the centre of mass prediction ($p=3$) are shown as dashed black line and solid red line, respectively.}
	
	\label{r1r2}
\end{figure}

To further characterise the coalescence preference, the angles of the merged droplet and the two parent droplets are used to quantify their respective change in position, as shown in Figure \ref{r1r2}A.  The ratio of the two angles is plotted against the size ratio of two parent droplets in Figure \ref{r1r2}B-C.  The experimental data (circle points) for the angle change are fitted (dashed black line) with the power law, $\Delta\theta_1 / \Delta\theta_2 \sim (r_1/r_2) ^{-p}$,  yielding an exponent $p$ $=$ 2.1 (SE= 0.042, R$^2$=0.71). The colorbar provided in Figure \ref{r1r2}B-C correspond to the larger droplet size ($r_2$) and the $Oh$ number, respectively. For this range, there is no apparent trend in $Oh$ in the deviation from the experimental fit. From the mapping of $r_2$ there appears to be increased scatter for the smaller droplet sizes when the ratio of droplet size is near unity. The source of outliers here is likely explained by surface inhomogeneities which will modify the pinning behaviour.  

Possible mechanisms for the coalescence preference may include the roles of centre of mass, surface energy release, or confinement. First, considering the centre of mass, if we assume there is no pinning, the final droplet position is determined by  $\theta_f = (m_1 \theta_1 + m_2 \theta_2) / (m_1 + m_2)$. The masses $m_{1,2}$ are obtained from the density and geometry of the droplets, assuming that the shape of the droplets is spherical cap and neglecting the shape alternation by the microcap.  The droplet volume can be described as $V = \frac{1}{6} \pi h(3r^2 + h^2)$ where $r$ and $h$ are the base radius of the droplet and height, respectively. While the base radius is directly measured, the height is estimated to be $h = r(1 - \cos{\phi_c})/\sin{\phi_c}$ where $\phi_c$ is the contact angle of the droplet. Assuming the contact angle is constant with droplet size, the volume then goes as $V$ $\sim$ $r^3$. The shift of each droplet is then predicted by $\Delta \theta_{n} = \theta_f$ - $\theta_{n}$. Following this approach, $\Delta\theta_{1}$ $\cdot$ ${r_1}^3$ = $\Delta\theta_{2}$ $\cdot$ ${r_2}^3$, or $\Delta\theta_{1}/$ $\Delta\theta_{2} \sim$ ${(r_1 /r_2)}^{-3}$, thus we expect $p$ = 3 if the coalescence occurs according to the centre of mass, indicated by the solid red line in Figure \ref{r1r2}B-C. Evidently, there is a clear difference between the experimental data and the centre of mass prediction.

Following the same approach as Weon and Je \cite{Je2012} who studied both bubble and droplet coalescence, if the position is proportional to kinetic energy $k$ gained by surface area ($s$) reductions, one should arrive at a scaling exponent $p$ = 5.3; namely $\Delta\theta_1 / \Delta\theta_2 \sim (r_1/r_2)^{-5.3} \sim (k_1/k_2)^{-5.3}$, where $k_1$ $\sim$ $(s_f - s_1)/s_1$, $k_2$ $\sim$ $(s_f - s_2)/s_2$. This was shown to fit well for bubbles at an oil-water interface with $p$ $\approx$ 5.1 and to a lesser extent for water droplets in decalin with $p$ $\approx$ 4.3. Clearly this is not the case here, and arguably not expected, as $Oh$ $\sim$ 1 - 3 would indicate there is minimal surface energy release. 

For $Oh$ $\sim$ 1 - 3, the coalescence at early stages is expected to follow the inertially limited viscous regime (ILV) before viscous flow dominates at latter stages.\cite{regimes} In both regimes, the velocity of the back of the droplet (point furthest from the connecting bridge) can be theoretically described.\cite{regimes} By integration with respect to time the approximate displacement may further be derived. From this the expected displacement will scale with $r^{-3}$ and $r^{-1}$ for the ILV and viscous regimes respectively. Either regime alone cannot account for the exponent determined here. In both cases, the capillary force $F$ which drives the droplets together,  $F = 2\pi\gamma r_{min}$, where $r_{min}$ denotes the initial bridge neck radius, is balanced against the force required to move the droplet mass. For small surface droplets with a low contact angle, it is plausible that the energy required to move the droplet may be governed by contact angle hysteresis.\cite{coalescence_hysterysis} The energy to overcome this hurdle is dependent on the footprint area and will scale with $r^2$. Accordingly, if $F$ is balanced against this energy requirement, displacement would be expected to follow $\Delta\theta_{1}/$ $\Delta\theta_{2} \sim$ ${(r_1 /r_2)}^{-2}$, yielding an exponent of $p$ = 2, which is consistent with the experimental results.  Future experiments measuring the true contact area between the droplet and both the flat and the microcap substrate during growth are necessary to verify the relevant surface energy contributions during the various coalescence events. 

The exponent determined herein are also consistent with two recent reports studying the coalescence of bubbles. Recent work \cite{preference_dense} has demonstrated a scaling of $p$ $\sim$ 2.06 $\pm$ 0.33 for bubble coalescence. The key feature in that experiments was that the bubbles were confined with a volume packing of $\sim$ 40 \%. The small exponent $p$ was attributed to the dense packing, which limits the motion of the merged bubble. A numerical study also arrived at a scaling exponent $p = 2.08$.\cite{lbm_preference} In that case the size preference was attributed to unbalanced pressures which drives increased flow towards the larger parent bubble.  In comparison to the surface nanodroplets studied herein, the droplet/bubble size in the existing studies\cite{Je2012,preference_dense,lbm_preference} is considerably larger, on the order of 10s - 100s $\mu$m, yielding $Oh$ $\sim$ 0.01 - 0.06.  With the reduced scale for surface nanodroplets and higher $Oh$, it is reasonable that the mechanism for coalescence preference becomes surface dominated, as described above. The results herein invites future research towards the coalescence of small droplets when $Oh$ < 1.

\subsection{Universality in symmetrical arrangement of droplets}
\begin{figure*}[htp]
	\includegraphics[trim={0 8.5cm 0 0}, clip,width=2\columnwidth]{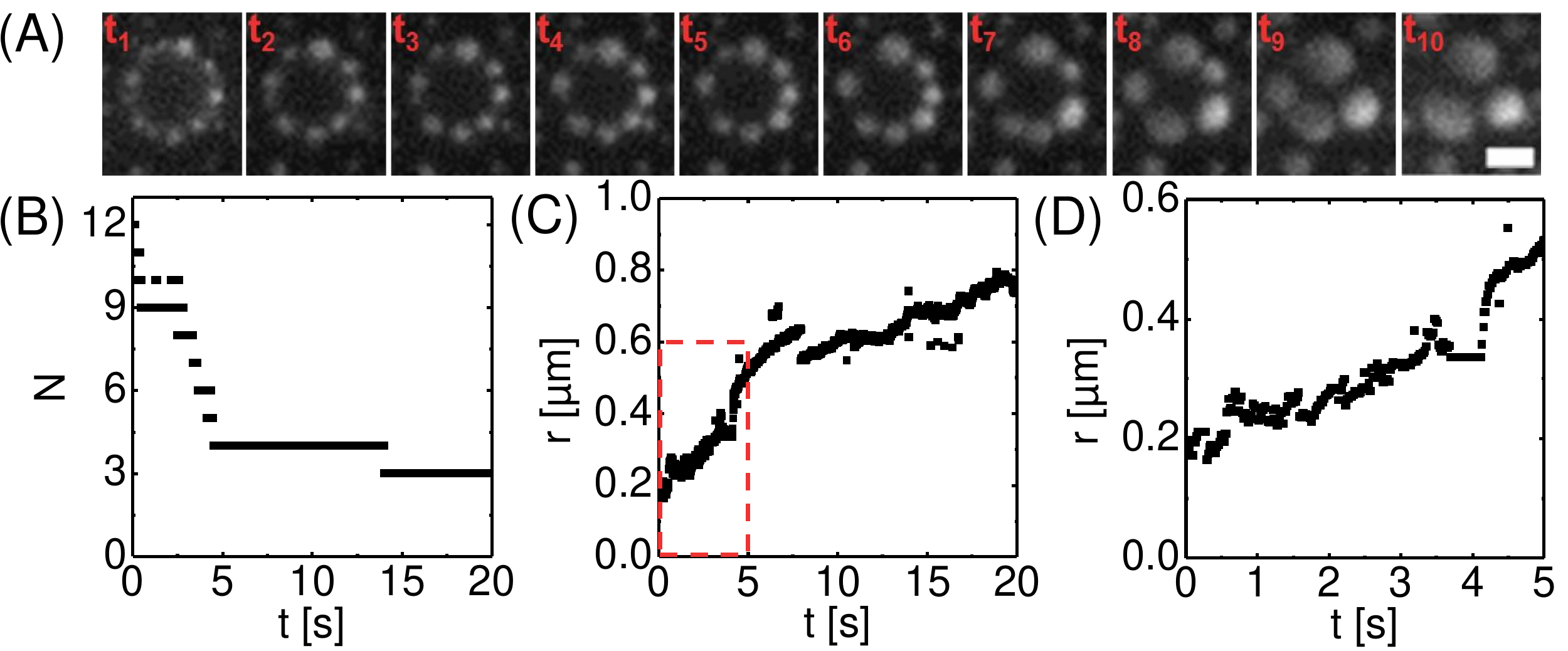}
	\caption{(A) Snapshot the droplet growth around a 3.46 $\mu$m cap. As the droplets grow, the number of droplets is reducing and the angle between the droplets is increasing. The corresponding times $t_{(1-10)}$ are $\sim$ 0 s, 1.3 s, 2.7, 3.1 s, 3.3 s, 3.6 s, 3.8 s, 4.1s, 4.6 s and 14 s. Scale bar = 2 $\mu$m. (B) Corresponding plot of number of droplets vs time and (C-D). The average droplet radius vs time (over 20 s and 5 s respectively) for HDODA droplet formation around a microcap with a base diameter of 3.46 $\mu$m. D is a zoom-in into the framed area of C. The increased resolution highlights the jumps in the average droplet radius, corresponding to the coalescence events. }
	\label{general_series2}
	\centering
\end{figure*}

Apart from around the rim, there are also incidents of droplet coalescence in other locations. Example coalescence events of the rim-flat region are shown in the left column Figure S2. A similar, yet less prevalent event is the coalescence of droplets atop of the cap and at the rim, examples of which are shown in the right column of Figure S2. In both cases, the droplets initially at the rim of the cap exert an extensively larger pinning force during coalescence, dominating the final position of the droplet as shown in Figure S2C. From these snapshots, it becomes evident that the droplets while being mobile atop and around the cap, will remain pinned to the cap even during coalescence to larger droplets on the flat substrate.

Consequently, the nanodroplets in our experiments are confined at the rim with freedom in azimuthal direction, but not over the geometrical bump of the microcap. The coalescence results in the droplets being continually re-positioned around the microcap according to the preference.  Namely, the coalescence between similar sized droplets will simultaneously result in movement away (angle-wise) from other droplets around the cap, increasing the angle between remaining droplets. Similarly, coalescence between dissimilar pairs will increase the angle by minor movement and removing the nearest neighbour. Consequently, as the number of droplets decreases, the average angle between the centroids of the remaining droplets increases.  At various time intervals these adjustments provoke a symmetrical rearrangement around the cap.

Our recent experiments showed that nanodroplets formed around the rim of a microcap surrounding by flat surface area exhibit symmetrical arrangements\cite{Peng2015}. The microcap size, droplet size and number follows a universal correlation, namely, the ratio between the cap radius and the droplet size is approximately 0.64 times the droplet number per cap. We explained this relation with a simple geometric argument\cite{Peng2015}. We now verify this correlation for a much larger range of microcap and droplets. Figure \ref{general_series2} shows the droplet growth around a microcap that is 3.46 $\mu$m in diameter, comparable to the typical cap size in our previous work \cite{Peng2015}. Based on the processed images, we are able to obtain time-dependant droplet growth with great detail. The plots in Figure \ref{general_series2}B-D are the number of droplets and the averaged droplet radius as a function of time. At the beginning, there are $N=12$ droplets with radii of $\sim 185$ nm.  The droplet number decreased rapidly to $N=4$ within 5 s. It then took $\sim$ 10 s to $N=3$ consequently. At each moment of coalescence, there are clear jumps in the averaged radius of the droplets. The sudden increases in mean droplet radius over the initial 5 s are shown in Figure \ref{general_series2}D. The three droplets of the final image in Figure \ref{general_series2}A remain symmetric around the rim for a long time. The angle in between them is about $120^o$. 


\subsection{Universal relation between droplet size and mutual angle}
\begin{figure}[!ht]
	\includegraphics[trim={0 0.5cm 2cm 0}, clip,width=1\columnwidth]{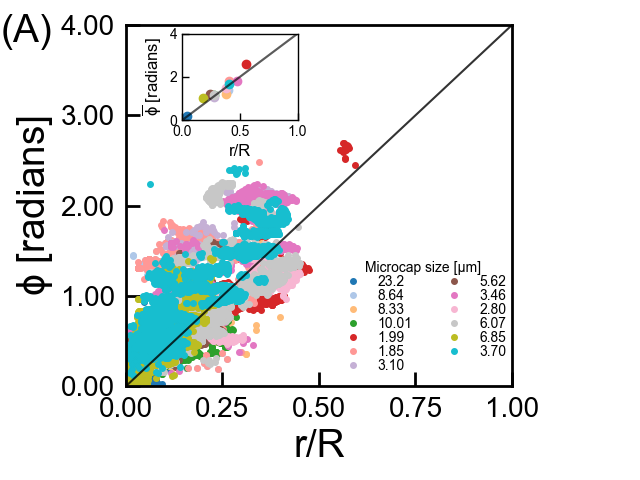}
		\includegraphics[trim={0 0.5cm 2cm 0}, clip,width=1\columnwidth]{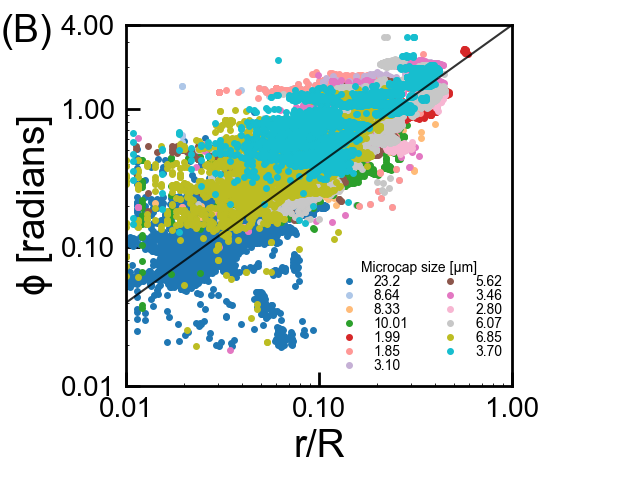}
	\caption{Measured angles ${\phi}$ between droplets vs $r/R$, on a linear scale (a) and on a double-logarithmic scale (b).
	The straight line shows relation (\ref{equation1}). 
	The microcap sizes are 1.85 $\mu$m (peach), 1.99 $\mu$m (red), 2.80 $\mu$m (pale pink), 3.10 $\mu$m (lavender), 3.46 $\mu$m (purple), 3.70 $\mu$m (teal), 5.62 $\mu$m (brown), 6.07$\mu$m (grey), 6.85 $\mu$m (olive)  8.33 $\mu$m (orange), 8.64 $\mu$m (light blue), 10.01 $\mu$m (green), 23.20 $\mu$m (blue). 
	The inset in (a) shows the final observed mean angle $\bar \phi$  for each microcap.}
	\label{symmetry}
	\centering
\end{figure}

But what determines the relation between the typical droplet size $r$ and the angle ${\phi}$  between them? Figure \ref{symmetry} shows these angles between neighbouring droplets against the size ratio 
$r/R$ for microcap sizes $R$  ranging from 1.85 $\mu$m - 23.2 $\mu$m (differentiated by colour) throughout the growth process,
on a linear scale (a) and on a double logarithmic scale (b).
 The inset in (a) shows the final average angle $\bar{\phi}$ vs $r/R$ for each microcap.   For each of the microcaps it can be seen that the angles $\phi$ between the droplets have a relatively sharp distribution and
roughly obey  
\begin{equation}
{\phi}  \approx \frac{4r}{R}.
\label{equation1}
\end{equation}
This then obviously also holds for the average angle $\bar \phi $ (for which by definition it holds
$\bar\phi \equiv  {2\pi / N}$),   i.e., 
\begin{equation}
\bar{\phi} \equiv \frac{2\pi}{N} \approx \frac{4r}{R} . 
\label{equation2}
\end{equation}
 Equation \ref{equation2}  universally holds  for all studied droplets in the 
  large range of cap sizes from $R$ = 2 $\mu$m up to   20 $\mu$m and for all observed droplet sizes from the  
  the very small ones with 
   $r \approx$ 200 nm  (low values of $r/R$) 
   up to the largest ones in the end which have grown up to  several micrometers (higher values of $r/R$).
   As explained in ref.\ \cite{Peng2015}, the interpretation of equation (\ref{equation1}) is that each droplet of radius $r$ together with 
   its diffusive  boundary layers requires a lateral space
   of $4r$, namely about $2r$ for its diameter and $1r$ for each diffusive  boundary layer on each side. 
As the droplets are all roughly of the same size, the total lateral space is $N \times 4r$, which then should 
correspond to the circumference $2\pi R$ of the microcap, i.e., equation (\ref{equation1}). 

One striking feature of figure \ref{symmetry}A is that in certain cases the angles seem to be discretized to certain values, while $r/R$
can take any value. The reason is pinning: During the diffusive growth process the droplets grow in diameter and also in 
distance of their center to the center of the cap around which they are sitting. But if no coalescence events take place, then
$\phi$ between the droplets remains fixed and all what happens is that new  data points in figure \ref{symmetry}A 
are added on the right (larger $r/R$), while the angle between the growing droplets remains exactly the same. This only changes
once the droplets merge.

Relation (\ref{equation1}) has also been shown to be true for multiple oils.  The physical properties for each oil are listed in Table 1. The interfacial tension of HDODA and 1-octanol with water is 35 $mN/m$ and 8.52  $mN/m$, respectively. Regardless of such difference, 1-octanol droplets also self-organize around the microcap rim in the same manner as HDODA droplets.  Results for 1-octanol droplets formed around a 8.33 $\mu$m, 8.64 $\mu$m and 10.01 $\mu$m microcap are included in Figure \ref{symmetry}. The results show  that 
relation (\ref{equation1}) between 
 the angle between the droplets and the radius ratio between droplet and microcap is also universal for different droplet liquids. 


\subsection{No sign for Ostwald ripening driven self-organization } 

In our previous work, the symmetrical arrangement and the relation between the size and number of droplets around the cap was attributed to Ostwald ripening: the enhanced growth of larger droplets is at expense of smaller droplets in an oversaturated environment \cite{Peng2015}.  The centre position of the larger droplet shifts towards the consumed smaller droplet, due to the concentration field biased by droplet dissolution. Indeed, in the case of microbubbles on a microwell array, we were able to visualise dissolving bubbles and symmetrical arrangement at certain stage of bubble growth \cite{Collectivebubble}.  However, from Figure \ref{individualgrowth} we observed no incidents of droplets shrinking. Moreover, our TIRF measurements show that the coalescence clearly leads to the shift in the droplet position, leading to nearly symmetrical arrangements of droplets around the microcap with time.
 
Considering that dissolution of small bubbles was observed in collective growth, one may argue that the interfacial tension of the droplets may be important for Ostwald ripening to occur, which may be spoilt by the fluorescent dye (nile-red) in TIRF measurements. To exclude this, we measured the contact angle of two types of drops in water.  After doping the dye, the contact angle of a HDODA drop in water diminished from $\approx 18^\circ$ to $\approx 14^\circ$, while 1-octanol drop remained at $\approx 54^\circ$. Nonetheless, we observed similar self-organization of these two types of surface nanodroplets. Moreover, using only a water soluble dye (rhodamine 6G),
 we found the same results: no small droplet dissolved as the droplets arrange around the rim.   Therefore, it is clear that here the droplet arrangement can not be attributed to Ostwald ripening, as hypothesised in our previous work for the symmetrical conditions of that paper.

\subsection{A dynamical model for coalescence driven self-organization}
 
\begin{figure}[!ht]
	\centering
	\includegraphics[trim={0 1.3cm 0 0}, clip,width=1\columnwidth]{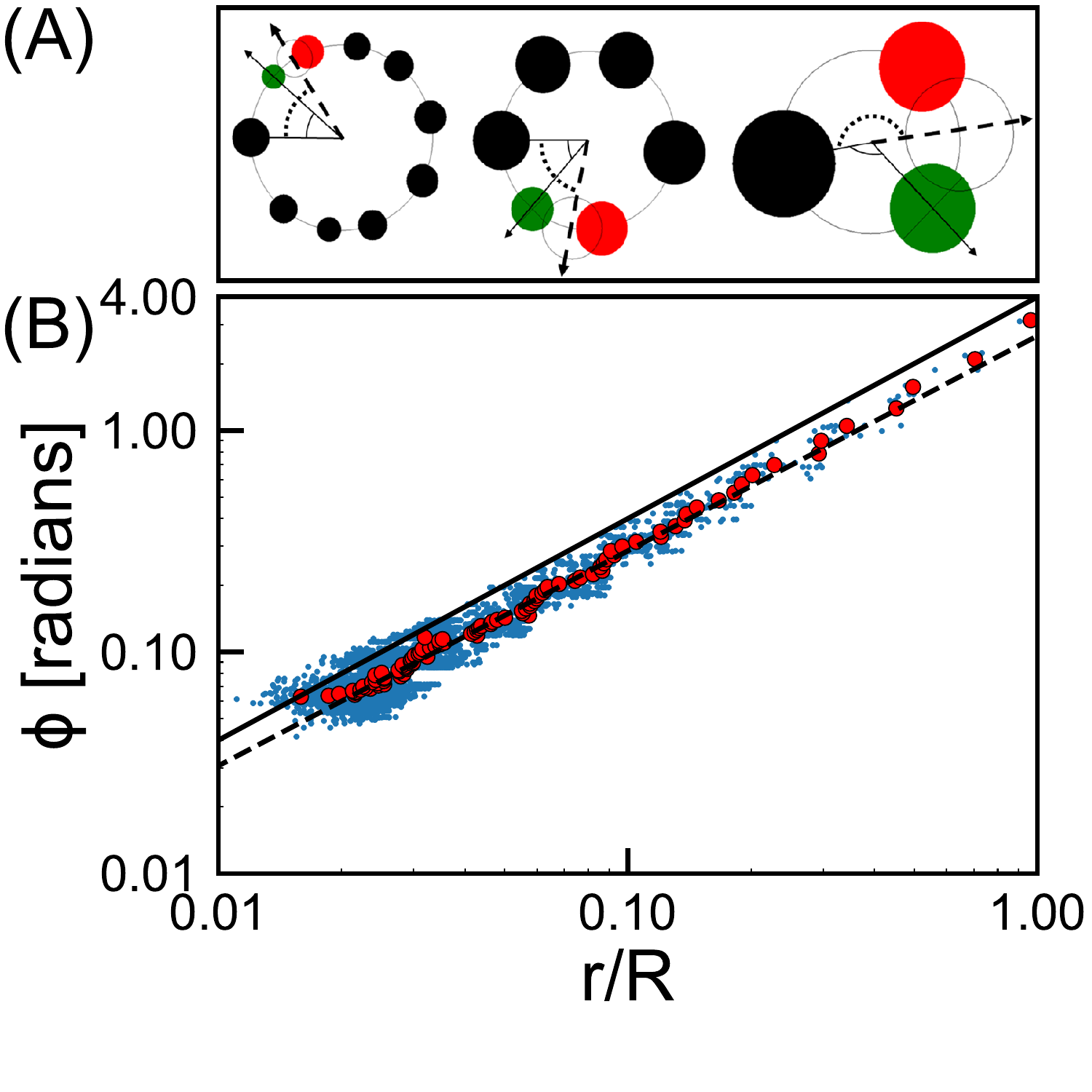}
	\caption{(A) Representation of  the coalescence model when $N$ = 9, 6, 3, respectively. In each sketch the two droplets predicted to coalesce are shown in green (smaller droplet) and red (larger droplet). The final predicted position is shown by the unfilled black circle. The respective angles $\phi$ between the droplets before and after the coalescence process are
	also shown. 
	(B) 
	Angle $\phi$ (light blue) and average angle $\bar \phi$ (red) between droplets vs $r/R$ as revealing from the model.  
	A power law fit to the data gives $\bar \phi = 2.67 (r/R)^{1.002\pm 0.005}$. 
	Equation (1) with $\bar \phi = 4 r/R$ 
	 is given as black line.}

	\label{model}
	\end{figure}

How then does  the distribution in angles between neighbouring droplets arise? 
 Here we suggest an iterative dynamical model which describes the evolution of the angles between
 neighbouring droplets. 
 \begin{enumerate} 
 \item
 We start off with $N$ 
  droplets  with an angular distance of   ${\phi} = 2\pi/N$ and an  initial radius obeying 
  $ 4r/R  = \phi = 2\pi /N $. We then slightly broadened these initial distributions in angle and radius, of course
  obeying the geometrical constrains. 
  \item 
  The growth of the droplets in time  is given by diffusion out of the oversaturated solution. Neglecting the diffusive
   interaction
  between the droplets (which we explored in ref.\ \cite{zhu2018-softmatter}) as this dynamical 
  model focuses on coalescence and not on Ostwald ripening effects, we estimate the diffusive growth
  as $\dot m \propto r^2 \dot r \propto r^2 \partial_r c \propto r$. The last relation holds 
  as the thickness of the diffusive boundary
  layer roughly equals $r$, i.e., $\partial_r c \approx (c_\infty - c_s) /r$, where the nominator is the concentration 
  difference between the droplet-water interface and the imposed oversaturation $c_\infty > c_s$.
   Thus $\dot r \propto 1/r$ or $(r(t))^2 - r_0^2 \propto
  t- t_0$ or $r(t) = \sqrt{r_0^2 + D^*(t-t_0)}$, where $D^*$ is the renormalized diffusion constant, whose absolute
  value is irrelevant here, as it only renormalizes time. 
  \item 
  We calculate the minimal time when two of the growing droplets touch. Therefore for each pair of neighboring
  droplets with angle $\phi$ between them and radius $r_{1,0}$ and $r_{2,0}$
  we calculate the distance $d_{1,2}(t) = \phi  R - r_{1}(t)  - r_{2}(t)  = 
  \phi R -   \sqrt{r_{1,0}^2 + D^* (t-t_0)} -  \sqrt{r_{2,0}^2 + D^* (t-t_0)}$ and identify the time $t_t$ when the first pair 
  touches, $d_{1,2} (t_t) = 0$. 
  \item
  That pair is then merged, according to the following coalescence rule: The two coalescing drops are removed
  and the position of the new coalesced drop is determined by the scaling law we found, namely, $\Delta\theta_{1}/$ $\Delta\theta_{2} \sim$ ${(r_1 /r_2)}^{-2}$ for the ratio of the relative shifts of the angular position. 
  The  volume of the droplets is additive, corresponding to $r_{new} = (r_1^3 + r_2^3)^{1/3}$. All other droplets keep
  the radius they have at $t_t$. After this merging step, the number of droplets is thus reduced by 1, from $N$
  to $N-1$. The merging process is sketched in figure \ref{model}A.
  \item We continue with step 2 and grow the droplets up to the next coalescence events.
  \item The coalescence process is stopped at $N=1$. Obviously, once $N$ become very small, the 
  employed geometric approximations of this calculations will become worse, but will still give the rough trend. 
 \end{enumerate} 
The numerical results from this model are presented in Figure \ref{model}B. 
As one can see, the overall features of $\phi \propto r/R$ is very well represented by this simple dynamical model,
though with 2.67  the prefactor of the best fit  is a little bit smaller than the prefactor 4 of equation (1),
as in our model the droplets only merge when touching and obviously the diffusive boundary layer is not 
properly represented 
in this simple model.  --
A movie of the growth and merging process of the droplets according to this model 
 is available in the supplementary materials.

Figure \ref{model2} shows the distribution of angles as function of the number $N$ of droplets and compares them to 
the experimental distribution, which is somewhat wider, reflecting more irregularities and pinning in the experiments 
as compared to our idealized model.

 \begin{figure}[!ht]
	\centering
	\includegraphics[trim={0 0.1cm 0 0}, clip,width=1\columnwidth]{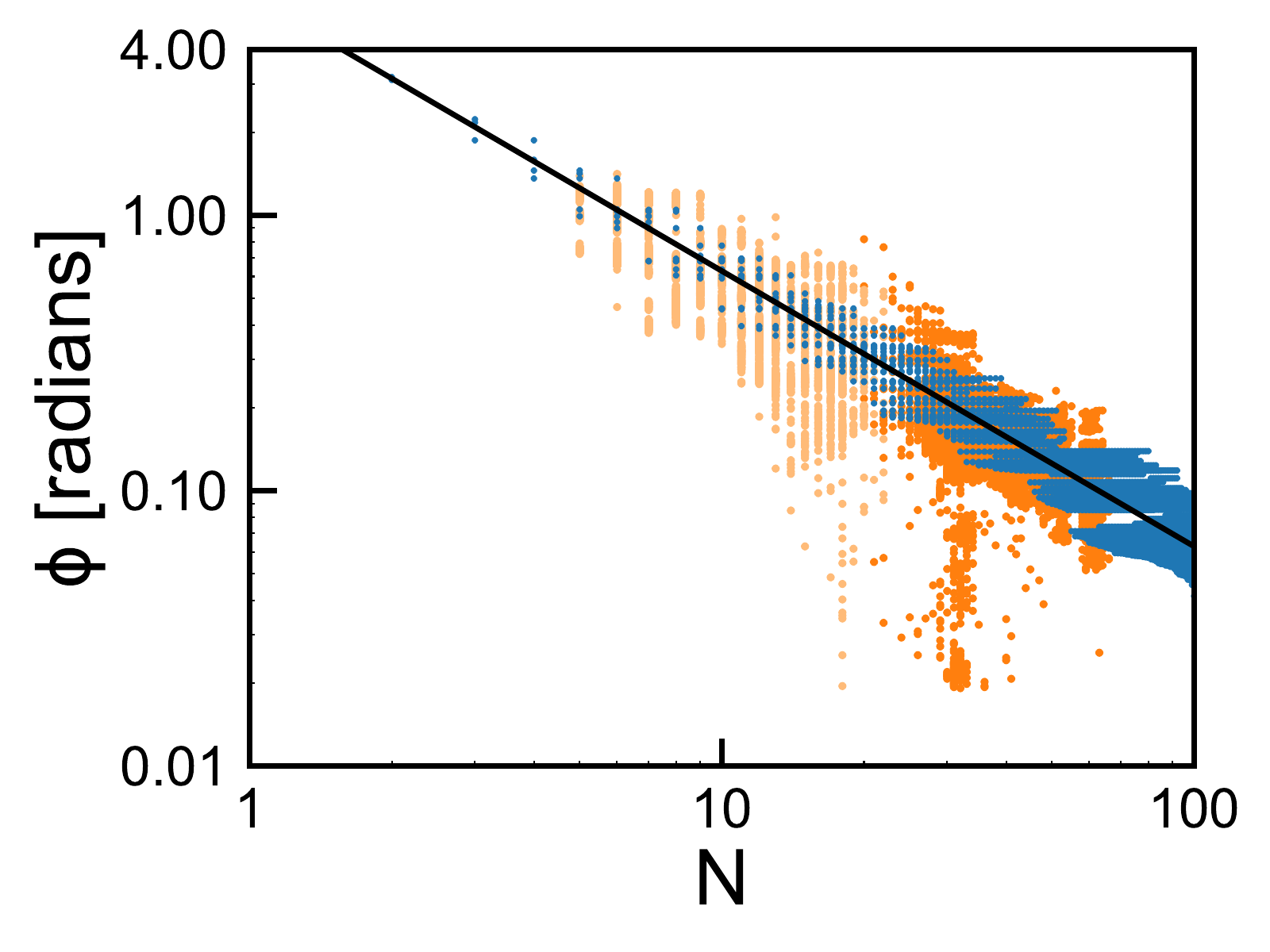}
\caption{
Distribution of angles as function of the number $N$ of droplets around the rim in the experiments (6.85 $\mu$m microcap - light orange \& 23.2 $\mu$m microcap - dark orange) and in the model (blue), in a log-log representation. The black curve is $2\pi/N$, which is the mean angle by definition.}
	\label{model2}
	\end{figure}


\section{Conclusion}
Growth and coalescence of surface nanodroplets by solvent exchange can be effectively captured \textit{in-situ} by the spatial and temporal resolution provided by TIRF. The growth behaviour around a microcap for nanodroplets demonstrates a self-ordered symmetry across various microcap and droplet sizes and droplet liquids. This self-organization behaviour is attributed to coalescence events. From studying the coalescence events it is clear that the final position of the droplets shows coalescence preference determined by droplet size inequality, following a scaling law with the size ratio of parent droplets. Two droplets of similar size, effectively will shift to meet each other at their centre location. As small droplets merge with larger droplets, the position of the merged droplet is located nearer to the larger droplets initial location. In both cases the angle between the remaining droplets around the microcap rim is obviously increased.  It was also revealed that droplets are strongly pinned to the microcap rim. During coalescence between droplets on and external to the rim location, the final droplet will be collected to the rim. Based on these observations, we built a simple dynamical model for the diffusive growth and the merging of the droplets around the rim which shows results consistent with the measured data. 

The findings in this work may be valuable for the future use of surface droplets as precursors in fabrication of a range of surface microstructures, including porous materials\cite{Bunz2006}, microlens for enhanced light diffraction and asymmetrical 3D surface structures with desirable wetting properties. Controlling size and spatial arrangement of droplets is essential to these applications.

\section*{Conflict of interest}
There are no conflicts to declare.

\section*{Acknowledgements}
We thank Dennis van Gils for programming the iterative model. 
X.H.Z. acknowledges the support from the Australian Research Council (FT120100473,LP140100594). We also acknowledge the RMIT MicroNano Research Facility for providing access to equipment and resources. This work was
supported by the Netherlands Center for Multiscale Catalytic Energy Conversion (MCEC), an NWO Gravitation programme funded by the Ministry of Education, Culture and Science of the government of the Netherlands.



\balance


\bibliography{literature} 
\bibliographystyle{rsc} 

\end{document}